# Robust Ferroelectricity in Silicon Dioxide upon Intercalation of Ammonia

Yaxin Gao[1,2], Menghao Wu[2*], Jun-Ming Liu[3]

[1]School of Physics and Mechanical Electrical & Engineering, Hubei University of Education, Wuhan, Hubei 430205, China.

[2]School of Physics, Huazhong University of Science and Technology, Wuhan, Hubei 430074, China.

[3]Laboratory of Solid State Microstructures, Nanjing University, Nanjing, Jiangsu 210093, China.

**Abstract**  The nanoelectronic applications of current ferroelectrics have been greatly impeded by their incompatibility with silicon. In this paper we propose a way to induce ferroelectricity in silicon dioxide ($SiO_2$), which is still the most widely used dielectric material in silicon-based chips. We show first-principles evidence that the intercalation of $NH_3$ molecules into crystalline $SiO_2$ is exothermic ($\Delta E$=-0.327 eV/molecule), where $NH_3$ molecules form quasi-bonds with $SiO_2$, giving rise to large and robust polarizations. In general, such polarization can be reversed via the reformation of N-Si bondings, which is multiaxial so vertical ferroelectricity may emerge in their thin-films of any facets. When the applied external electric field is large enough, however, the system may exhibit unconventional quantized ferroelectricity of unprecedented magnitude, where $NH_3$ may migrate for multiple lattice constants like mobile ions in ion conductors. Compared with ion conductors with charged mobile ions and ion vacancies that may lead to current leakage, herein the intercalated systems can be denoted as "neutral ion conductors" where both pristine $SiO_2$ and $SiO_2$ filled with $NH_3$ are insulating. Similar ferroelectricity may exist in various $SiO_2$ crystalline polymorphs, its amorphous phase, and other porous structures intercalated by $NH_3$. Our findings may not only resolve the bottleneck issues for the compatibility of ferroelectrics and silicon, but also develop unconventional mechanisms of ferroelectricity.

Introduction

Ferroelectrics with electrically switchable polarization have attracted consistent and increasing research interest due to fascinating physical phenomena and great potential in non-volatile memory device applications.[1] Currently, the most technologically important ferroelectrics are still perovskites,[2] while the synthesis of ferroelectric perovskite films on silicon are impeded by chemical incompatibility and the high temperatures required for epitaxial growth. In recent years, ferroelectric $HfO_2$ has emerged as a promising material as they enable low-temperature synthesis and conformal growth on silicon,[3,4] providing a solution to the issues that have hindered the adoption of perovskite-based ferroelectrics in metal-oxide-semiconductors. Meanwhile its ferroelectricity stems from the metastable nonsymmetric orthorhombic phase, which is stabilized by surface energy effects and element doping.[5]

In this paper we propose a way to induce ferroelectricity in silicon dioxide ($SiO_2$), which is still the most widely used dielectric material used in the microelectronics field due to its excellent dielectric strength, large band gap and low interface defect density with Si. Our previous study[6] shows that silicon surface may become ferroelectric upon ligand passivation, while such functionalization cannot be applied to the bulk phase where the inner space cannot accommodate a ligand. We note that the inner hollow space of $SiO_2$ is much larger, which may accommodate a small ligand inducing symmetry breaking. To maintain the insulating properties of $SiO_2$, the candidates for intercalation should be neutral molecules. We show first-principles evidence that robust ferroelectricity can be induced in $SiO_2$ via the intercalation of ammonia molecules, which form quasi-bonds with $SiO_2$ as well as large robust polarization, resolving the issues for the compatibility of ferroelectrics and silicon. Such ferroelectricity may exhibit unconventional multimodes of ion displacements, including long migration like in ion conductors that give rise to quantized polarizations.[7-13]

**Fig 1: The transformation of $SiO_2$ upon ammonia intercalation.**

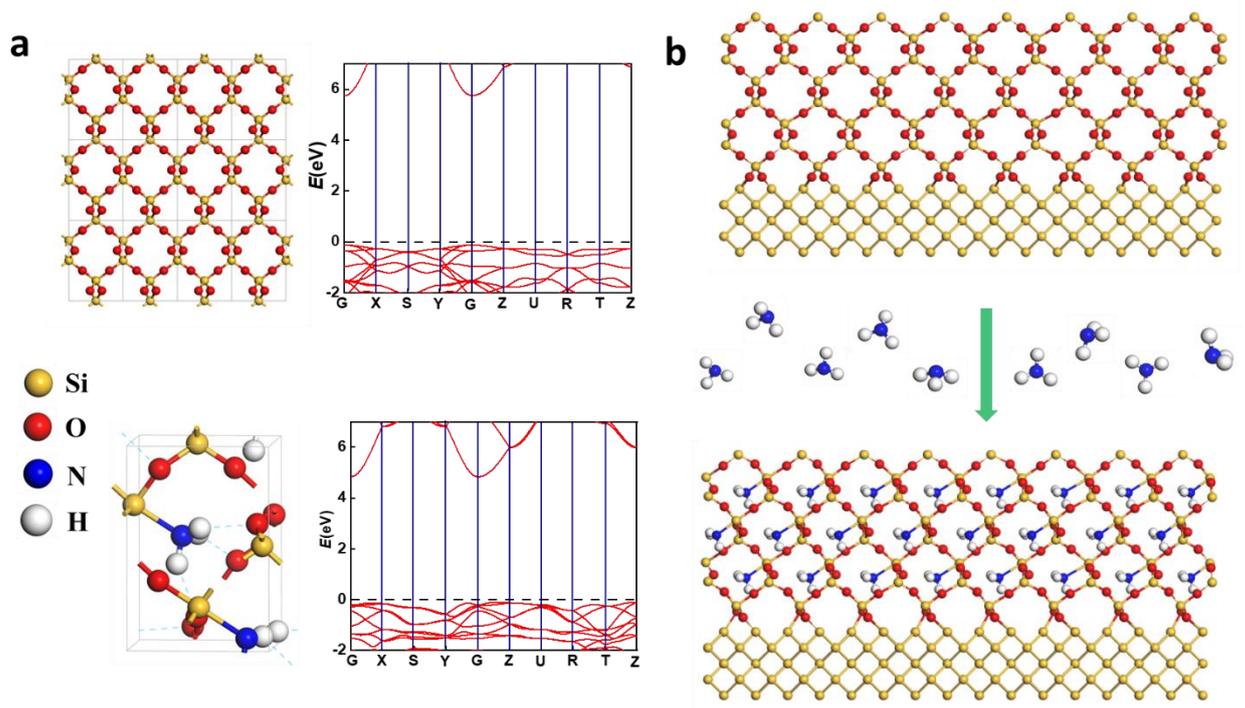

**a** Geometric structures and bandstructures of pristine SiO$_2$ crystal and SiO$_2$ intercalated by NH$_3$. **b** The formation of ferroelectric thin-film on silicon upon the diffusion of NH$_3$ into SiO$_2$ layer.

## Results

Among the various crystalline polymorphs of SiO$_2$, first we select β-cristobalite phase that was predicted to be the ground state (33.43 meV/f.u. lower in energy compared with quartz phase in previous calculations[14]) as a paradigmatic case. According to our calculations, when a NH$_3$ molecule is placed inside crystalline SiO$_2$, its N atom will be inclined to form a quasi-bond with a Si atom (with a Si-N bond length of 2.04 Å, notably larger compared with 1.63 Å for Si-O bonds in SiO$_2$). Based on Hirshfeld charge analysis, each quasi-bonded NH$_3$ carries a positive charge of 0.14 |e|, revealing a considerable charge transfer between SiO$_2$ and NH$_3$. For SiO$_2$ fully intercalated by NH$_3$, which may denoted as NH$_3^{2\delta+}$(SiO$_2^{\delta-}$)$_2$, its lattice is slightly enlarged and distorted (a=5.35 Å, b=5.06 Å, c=7.35 Å, α=85.2°, β=90°, γ=90°) compared with pristine SiO$_2$ (a=4.92 Å, b=4.92 Å, c=7.37 Å, α=90°, β=90°, γ=90°). Bandstructures in Fig. 1a reveals that SiO$_2$ remains insulating upon the intercalation of NH$_3$, making it still a good dielectric material. The binding energy $\Delta E = E(NH_3Si_2O_4) - E(NH_3$

molecule)- 2$E$(SiO$_2$ crystal) is estimated to be -0.327 eV per NH$_3$ molecule, where the negative value favors its formation. It might be synthesized in the ammonia gas (see Fig. 1b), where the diffusion NH$_3$ into SiO$_2$ can be further facilitated by applying high pressure and high temperature.

**Fig. 2: Configurations of different anchoring sites for NH$_3$ in SiO$_2$ and the transition pathways between them.**

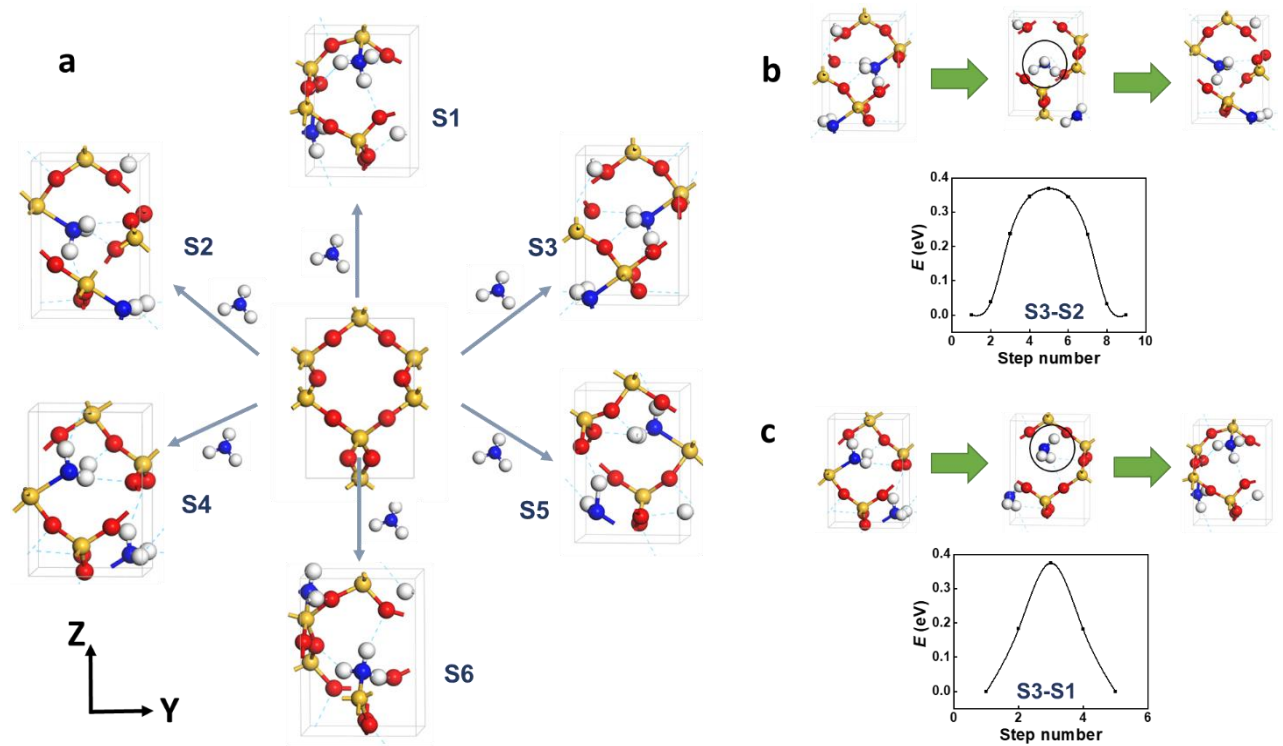

**a** 6 equivalent configurations for SiO$_2$ fully intercalated by NH$_3$. **b** The S3-S2 and **c** S2-S1 transition pathway, where NH$_3$ in the transition states are not bonded to Si, as marked by the black circles.

There are 6 equivalent binding sites for each N in a cavity, and the 6 equivalent configurations denoted as S1-S6 are displayed in Fig. 2(a). The migration of NH$_3$ to adjacent Si site involves breaking of the Si-N bond and reformation of a new one. For example, for the transition pathway S3-S2 and S2-S1 in Fig. 2b and c, the NH$_3$ molecules are not bonded to Si atoms in both transition states, which are 0.37 eV higher in energy per NH$_3$ compared with the ground state, so the barriers are both around 0.37 eV. Such migration of NH$_3$ changes the direction of Si-N bonding as well as the polarization, akin to the rotation of ligands in molecular ferroelectrics.[15,16] For example, the transition in Fig. 2b corresponds to the ferroelectric switching along –Y direction, and the evolution of polarization along the switching pathway reveals a large switchable polarization of 24.5 μC/cm$^2$, close to the value of BaTiO$_3$. Meanwhile such ferroelectricity is multiaxial, and vertical switchable polarization may emerge in their thin-films of any facets. The thermal stability of ferroelectricity is further revealed by ab initio molecular dynamics (AIMD) simulation in Fig. S1, where the directionality of Si-N bondings can be maintained at 700K.

**Fig. 3: Long displacement mode of NH$_3$ in SiO$_2$.**

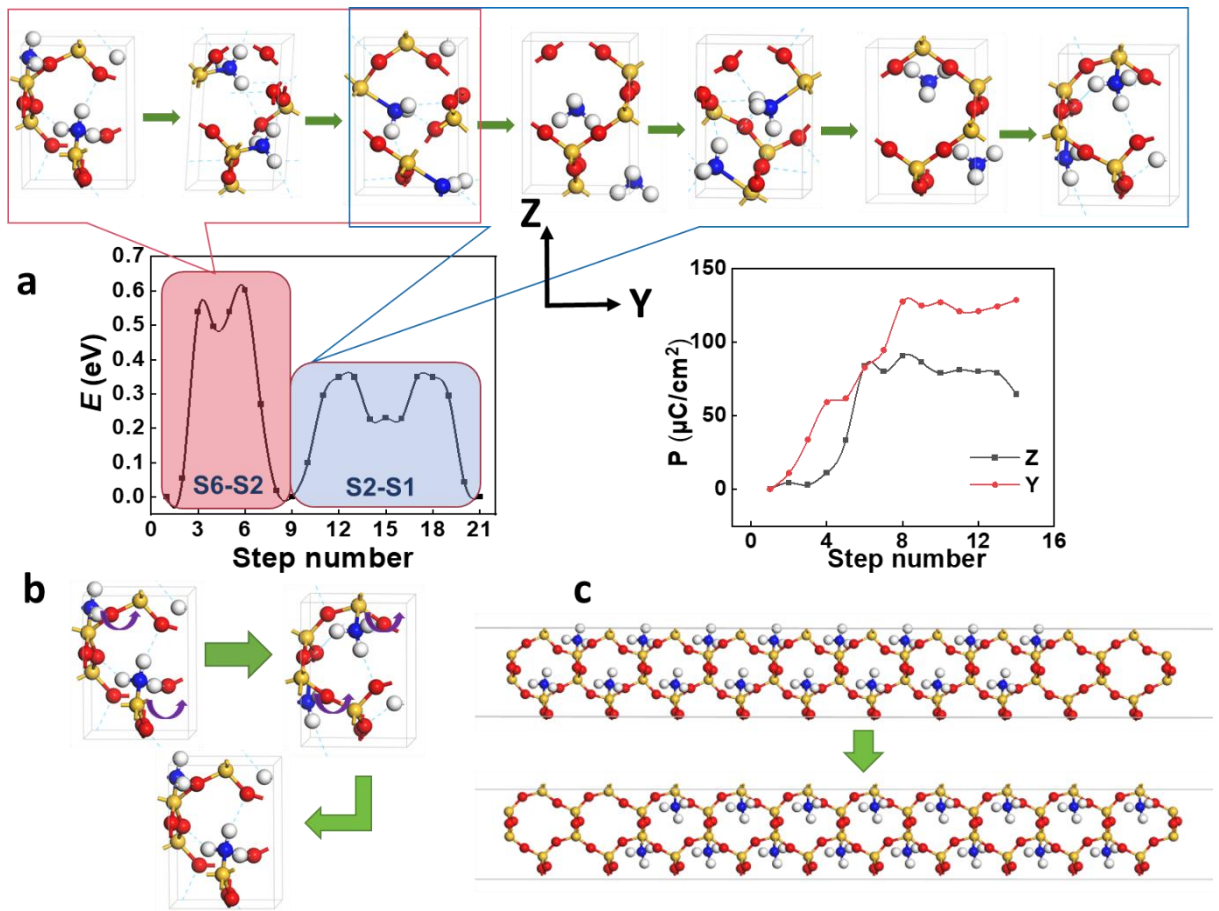

**a** The transition pathway of S6-S2-S1 and the corresponding polarization evolution, where NH$_3$ molecules are displaced by half lattice constant along –Y direction. **b** The migration of NH$_3$ molecules by a whole lattice constant along –Y direction can be achieved via S6-S1-S6 transition, and the barrier will be the same for migration of multiple lattice constants. **c** In SiO$_2$ thin film of 10 unit thickness with 80% filled by NH$_3$, all NH$_3$ can migrate for 1.5 lattice constants along –Y direction driven by electric field.

Another pathway to reverse the polarization involves the translation of $NH_3$ to the adjacent cavity without breaking the Si-N bond. As shown in Fig. 3a, the transition from configure S6 to S2 may also be realized via the rotation of Si-N bondings, and the switching barrier is estimated to be around 0.6 eV. Driving by an electric field along –Y direction, the $NH_3$ molecules may continue to migrate and form into configuration S1 via reforming new Si-N bonds, crossing a barrier of 0.37 eV (same as S3-S2 and S3-S1 in Fig. 2). In such S6-S2-S1 process, each $NH_3$ is displaced by half a lattice constant along –Y direction, while the polarization along –Z direction is also reversed. The evolution of polarization during the migration in Fig. 3a reveals switchable polarizations of **65.0 and 32.3 µC/cm²** respectively along –Y and –Z direction, where the polarization along –Y direction is quantized that corresponds to quanta $\Delta n = \Omega \Delta P / e|\mathbf{a}| = 3$ (where $\Omega$ is the unitcell volume and $|\mathbf{a}|$ is the lattice constant)[8].

For the migration of $NH_3$ along –Y direction, no matter for half lattice constant, a whole lattice constant or multiple lattice constants, the barriers should be the same based on the crystal symmetry. If an electric field applied along –Y direction is enough to concur the switching barrier of 0.37 eV in Fig. 2 but still not enough for the barrier of 0.6 eV in Fig. 3, the system will exhibit ferroelectricity that is switchable via the reformation of Si-N bonds shown in Fig. 2. When the electric field is large enough to concur the latter barrier, $NH_3$ may migrate for multiple unitcells (depending on the number of ion vacancies in each ion conduction channel) like mobile ions in ion conductors, giving rise to large quantized ferroelectricity.[7,8,17,18] For example, in $SiO_2$ thin-film of 10 unitcell thickness with 80% filled of $NH_3$, all $NH_3$ can migrate for 1.5 unitcell, as shown in Fig 3c, with a switchable polarization of **195 µC/cm²**. Theoretically, this value can be enhanced by two orders of magnitudes (~20000 µC/cm²) in a 2 µm thick $SiO_2$ thin-film with 95% filled of $NH_3$. Such migrations also facilitate the diffusion of $NH_3$ deep into $SiO_2$ for the synthesis shown in Fig. 1(b). Meanwhile compared with ion conductors where the ion vacancies are charged and may lead to current leakage, herein the intercalated systems are "neutral ion conductors" where both pristine $SiO_2$ and $SiO_2$ filled with $NH_3$ are insulating.

**Fig. 4: Similar ferroelectricity in other hollow systems intercalated by NH₃ or H₂O.**

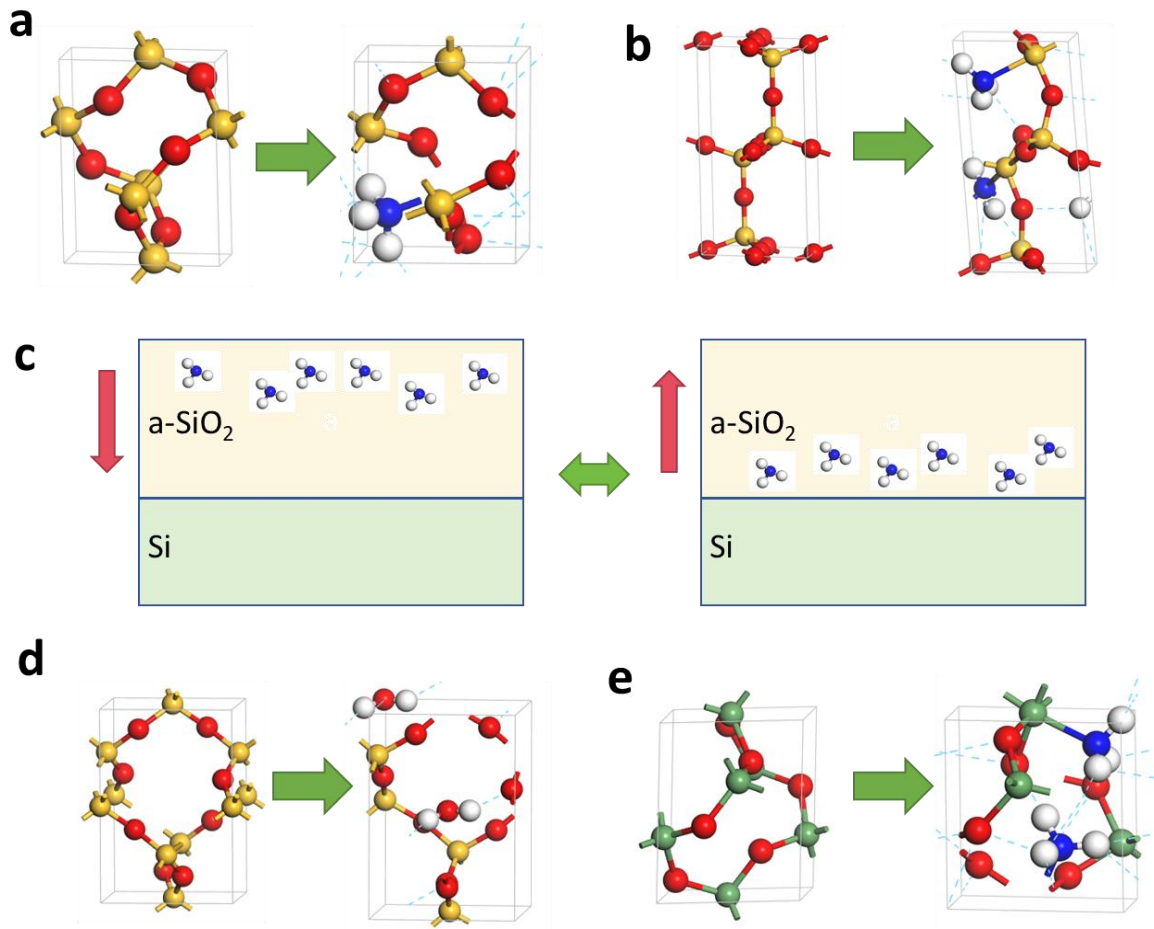

**a** α-quartz and **b** β-tridymite crystalline SiO₂ intercalated by NH₃. **c** In amorphous SiO₂ thin film partially filled by NH₃, vertical polarization (with directions marked by red arrows) may still be reversed via migration of NH₃ driven by gate voltage. **d** SiO₂ and **e** GeO₂ respectively intercalated by H₂O and NH₃.

**Discussion**

For well-known crystalline polymorphs of SiO₂ like quartz, tridymite, cristobalite, coesite, keatite,[14] they are all constructed from SiO₄ tetrahedra with small bond angle distortion, where each oxygen atom is shared between two tetrahedral. Si-N quasi-bonds are all likely to be formed when NH₃ molecules are placed in their hollow spaces, and similar ferroelectricity induced by such symmetry breaking may emerge. For example, the Si-N bond lengths are respectively 1.90 and 2.11 Å for α-quartz (with a high density of 2.61 g/cm³)[14] and β-tridymite

(with a low density of 2.00 g/cm$^3$)[14] crystalline $SiO_2$ intercalated by $NH_3$, as shown in Fig. 4a and b, which are both insulating with band gaps close to 5 eV (see Fig. S2). Compared with ground state β-cristobalite (with a density of 2.17 g/cm$^3$), the binding energy of $NH_3$ is stronger for β-tridymite ($\Delta E$=-0.41 eV/molecule), which becomes a positive value of 0.41 eV/molecule for α-quartz. It seems that lower density with larger hollow space favor such binding of $NH_3$, while smaller hollow space leads to larger repulsion. It is noteworthy that the directional ferroelectricity in Fig. 2 is likely to be absent in amorphous $SiO_2$ since the directionality of Si-N bondings will be in disorder, despite it may exist locally. However, such structure still maintains hollow spaces for $NH_3$, which should be able to migrate under an electric field, so the long displacement ferroelectricity in Fig. 3 may still exist, as shown in Fig. 4c.

The approach for inducing ferroelectricity via intercalation of small molecules may be applied to other systems. We have also investigated other molecules like $H_2O$ for intercalation, while it turns out that its binding to $SiO_2$ is much weaker ($\Delta E$ =-0.11 eV/molecule) without the formation of quasi-bonds (see Fig. 4d). For some other similar oxides like $GeO_2$ with larger inner space, the binding energy of $NH_3$ upon intercalation is larger (-0.46 eV/molecule), where two $NH_3$ molecules are inclined to form two quasi-bonds together with one Ge atom (see Fig. 4e). Similar properties upon intercalation may emerge in other silicon-based porous structures like zeolites.

In summary, we predict that giant robust ferroelectricity can be induced in $SiO_2$ upon the intercalation of $NH_3$ molecules that form quasi-bonds with $SiO_2$. Depending on the magnitude of electric field, the ferroelectricity can be either reversed via the reformation of N-Si bondings, or $NH_3$ may further migrate for multiple unitcells like mobile ions in ion conductors, giving rise to quantized polarization with unprecedented magnitude. Our prediction does not only propose a way of resolving bottleneck issues for practical applications, but also exploit new types of ferroelectricity. There is no doubt that such $SiO_2$–based ferroelectricity, if experimentally realized, will lead to revolution in both the industry of ferroelectric devices and the field of ferroelectric physics, and the key point for its realization is the synthesis of low-density $SiO_2$ structures that facilitate the diffusion of $NH_3$ inside.

## Methods

Our first-principles calculations were performed within the framework of density functional theory (DFT) implemented in the Vienna Ab initio Simulation Package (VASP 5.4.4) code[19,20]. The generalized gradient approximation (GGA) in the Perdew-Burke-Ernzerhof (PBE)[21] form for the exchange correlation potential, together with the projector augmented wave (PAW)[22] method, were applied. The kinetic energy cutoff was set to be 520 eV, and the Monkhorst−Pack *k*-meshed[23] were set to 7×7×5 in the Brillouin zone. The DFT-D2 functional of Grimme[24] was used to account for the van der Waals interactions. The geometric structures were fully relaxed until the energy and forces were converged to $10^{-6}$ eV and 0.01 eV·Å$^{-1}$, respectively. The Berry phase method[25] was employed to calculate ferroelectric polarizations, and the ferroelectric switching pathway was obtained by using a generalized solid-state nudged-elastic-band (SS-NEB) method[26].

## Data availability

Relevant data generated in this study are provided in the article and Supplementary Information. All raw data that support the findings of this study are available from the corresponding authors upon request.

## Acknowledgements

This work was supported by the Natural Science Foundation of Hubei Province (No. 2023AFB254), the Scientific and Technology Research Project of Hubei Provincial Education Department (No. B2023188), and the Scientific Research Foundation of Hubei University of Education for Talent Introduction (No. ESRC20230001).

## Author information

Authors and Affiliations

School of Physics and Mechanical Electrical & Engineering, Hubei University of Education, Wuhan, Hubei 430205, China

Yaxin Gao

School of Physics, Huazhong University of Science and Technology, Wuhan, Hubei 430074, China

Yaxin Gao and Menghao Wu

Laboratory of Solid State Microstructures, Nanjing University, Nanjing, Jiangsu 210093, China.

Jun-Ming Liu

Corresponding authors

Correspondence to Menghao Wu (wmh1987@hust.edu.cn)

## Ethics declarations

Competing interests

The authors declare no competing interests.

Supplementary information